\documentclass[12pt, letterpaper]{article}
\usepackage[english]{babel}
\usepackage{amsmath}
\usepackage{amssymb}
\usepackage{mathtools}
\usepackage{amsthm}
\usepackage{nicefrac}
\usepackage{pifont} 
\usepackage{microtype}
\usepackage{caption}
\usepackage{xcolor}
\usepackage{float}
\usepackage{subcaption}
\usepackage{nccmath}
\usepackage{soul}
\usepackage{appendix}
\usepackage{graphicx}
\usepackage{natbib}
\usepackage{tikz}
\usepackage{layouts}
\usepackage{xcolor}
\usepackage{graphicx}
\usepackage{titlesec}
\titlelabel{\thetitle.\quad}

\usepackage{hyperref}
\PassOptionsToPackage{linktocpage}{hyperref}
\hypersetup{
           breaklinks=true,   
           colorlinks=true,   
           linkcolor = blue,
            urlcolor  = blue,
            citecolor = blue,
            anchorcolor = blue,
        }
\usetikzlibrary{decorations.pathreplacing}
\usepackage[ruled, noend, noline, linesnumbered]{algorithm2e}
\usepackage[noend]{algpseudocode}
\algnewcommand{\Input}{\textbf{Input:} }
\algnewcommand{\Output}{\textbf{Output:} }
\algnewcommand{\Draw}{\textbf{draw} }

\usepackage{url}
\usepackage{setspace}
\newcommand{\blind}{0}
\usepackage{changes}
\usepackage{todonotes}

\begin{document}

\doublespacing

\if0\blind
{
  \title{\bf Grid Particle Gibbs with Ancestor Sampling for State-Space Models
}
  \date{July 2024}
  \author{Mary Llewellyn, Ruth King, V\'{i}ctor Elvira, Gordon Ross \\
  \normalsize{\it{School of Mathematics, University of Edinburgh}}}
  \maketitle
} 

\if1\blind
{
  \bigskip
  \bigskip
  \bigskip
  \begin{center}
    {\LARGE\bf }
\end{center}
  \medskip
} \fi

\begin{abstract}
    We consider the challenge of estimating the model parameters and latent states of general state-space models within a Bayesian framework. We extend the commonly applied particle Gibbs framework by proposing an efficient particle generation scheme for the latent states. The approach efficiently samples particles using an approximate hidden Markov model (HMM) representation of the general state-space model via a deterministic grid on the state space. We refer to the approach as the grid particle Gibbs with ancestor sampling algorithm. We discuss several computational and practical aspects of the algorithm in detail and highlight further computational adjustments that improve the efficiency of the algorithm. The efficiency of the approach is investigated via challenging regime-switching models, including a post-COVID tourism demand model, and we demonstrate substantial computational gains compared to previous particle Gibbs with ancestor sampling methods. \\ 

    \noindent \textbf{Keywords:} Bayesian inference, hidden Markov models, importance sampling, particle Gibbs with ancestor sampling.
\end{abstract}

\newpage 

\section{Introduction}
Discrete-time state-space models (SSMs) describe observed time series data, $y_{1:T}=(y_1, \dots, y_T)$, as dependent on an unobserved and continuously-valued latent process, $x_{1:T}=(x_1, \dots, x_T)$ \citep{Durbin_2012}. The latent states evolve over time according to a first-order Markov process, referred to as the latent state process. The observed data at each time point, $y_t$, are modeled in the observation process as a function of the current latent state(s). Each process has an associated set of static model parameters. We denote the set of all static parameters by $\theta$. We assume, initially, that the state and observation spaces at each time point are one-dimensional. Thus, an SSM can be written mathematically in terms of the two processes and static parameters as 
\begin{align}
    p(y_t \vert x_t, \theta)&, \quad \text{(observation process)} \nonumber \\
    p(x_t \vert x_{t-1}, \theta)&, \quad \text{(latent state process)} \label{eq:SSM}
\end{align} 
for time points $t=1, \dots, T$ where $p(x_1 \vert x_0, \theta)=p(x_1 \vert \theta)$ defines the initial state distribution. The two distinct processes of an SSM provide flexibility, leading to their application in a variety of fields, including ecology \citep{King_2014, AugerMethe_2021}, economics \citep{Koopman_2004}, and neuroscience \citep{Lin_2019}. However, inference of SSMs is often intractable outside of special cases when the SSM is linear and Gaussian or the state space is discrete \citep{Durbin_2012, Kalman_1960, Rabiner_1989}. Specifically, Bayesian inference of the latent states and model parameters of general SSMs, i.e., targeting the joint distribution $p(x_{1:T}, \theta \vert y_{1:T})$, can be challenging since the joint distribution often only admits a closed-form expression up to proportionality.

Markov chain Monte Carlo (MCMC) methods can be applied for inference targeting $p(x_{1:T}, \theta \vert y_{1:T})$ since this joint distribution generally admits a closed-form expression up to proportionality \citep{Tanner_1987, Newman_2022}. Apart from in special cases, for example when the SSM is linear and Gaussian \citep{Kalman_1960}, MCMC updates of the latent states involve simulating new values via some specified proposal distribution. However, sampling the latent states from a proposal distribution that accurately captures the distributional characteristics of the latent states to yield good MCMC mixing is often challenging since the latent state distribution is often complex \citep{Fruhwirth_2004, Borowska_2023, Llewellyn_2023}. Several approaches have been proposed to efficiently update the latent states, including Gaussian approximation methods \citep{Kristensen_2016, vanderMerwe_2004}. Such approaches can be applied efficiently when the SSM is well-approximated by Gaussian distributions but can be inefficient for general nonlinear or non-Gaussian SSMs \citep{Carter_1994}. Latent states can also be updated in lower-dimensional blocks, requiring lower-dimensional and simpler proposal distributions for each block \citep{Fearnhead_2011}. However, this often leads to poor mixing when the states are highly correlated \citep{Shephard_1997, King_2011}.

Particle Gibbs algorithms use sequential Monte Carlo (SMC) approximations to design efficient MCMC approaches for general SSMs. The original particle Gibbs algorithm \citep{Andrieu_2010} proposes latent states from a conditional SMC `particle' approximation to $p(x_{1:T} \vert y_{1:T}, \theta)$. The model parameters are then updated using standard, and typically simple, MCMC updates targeting $p(\theta \vert x_{1:T}, y_{1:T})$, resulting in an MCMC algorithm targeting $p(x_{1:T}, \theta \vert y_{1:T})$. However, the particle Gibbs algorithm is known to suffer from `sample impoverishment' in conditional SMC steps and can therefore require many particles and a high computational cost to achieve reasonable MCMC mixing and convergence \citep{Kantas_2014, Chopin_2015, Wigren_2019}. Consequently, several variants of the original particle Gibbs algorithm have since been proposed. In particular, the particle Gibbs with backward sampling \citep{Whiteley_2010, Lindsten_2013} and particle Gibbs with ancestor sampling (PGAS; \cite{Lindsten_2014}) algorithms can be particularly efficient \citep{Berntorp_2017, Nonejad_2015} but can still incur a high computational cost if there is high sample impoverishment \citep{Rainforth_2016, Llewellyn_2023}. 

\newpage 
We propose an approach to improve the efficiency of particle Gibbs algorithms, focusing on a novel and efficient solution to the SMC sample impoverishment problem. Our proposed method builds on the observation that the optimal theoretical approach to minimizing SMC sample impoverishment simulates particles directly from the conditional posterior distribution of the latent states \citep{Branchini_2021, Chopin_2020, Elvira_2019}. However, typically this conditional distribution is intractable for general SSMs. Previous approaches simulate particles in approximately high posterior regions \citep{Andrieu_2003, Donnet_2017, He_2022}. One approach is the auxiliary particle filter \citep{Pitt_1999, Pitt_2001, Carpenter_2000}, which samples particles from an approximation to the optimal importance distribution at each SMC recursion. While the auxiliary particle filter often reduces sample impoverishment, the computational cost of such approaches can accumulate quickly when used within particle Gibbs algorithms \citep{Elvira_2018}. 

The approach proposed in this paper, referred to as the \emph{grid particle Gibbs with ancestor sampling} (GPGAS) algorithm, uses coarse, deterministic (discrete-valued) hidden Markov model (HMM) approximations to direct SMC particles to regions of high posterior mass. The approach can substantially improve SMC sample impoverishment, leading to an SMC algorithm with many fewer particles (without loss of precision compared to alternative approaches) or more accurate approximations of the conditional latent state distribution (for the same number of particles). We use the HMM SMC algorithm within particle Gibbs with ancestor sampling (PGAS) steps to update the latent states conditional on the model parameters, and the model parameters are updated using standard Gibbs or Metropolis-within-Gibbs steps.

 We demonstrate the efficiency of the GPGAS algorithm by focusing on a class of models that remain challenging to fit: regime-switching SSMs. These models embed an additional latent state process allowing the observation and latent state transition models to change abruptly. However, despite their widespread use \citep{Haimerl_2023, Hamilton_1989, Liang-qun_2009}, current computational methods for fitting the latent states and model parameters of general regime-switching SSMs can be inefficient due to the abrupt changes in the state process. We investigate the performance of the proposed GPGAS algorithm when applied to such models, including a challenging real-data case study focusing on tourism demand recovery in Edinburgh. The rest of the paper is structured as follows. In Section \ref{background}, we introduce the particle Gibbs and PGAS algorithms and motivate the proposed GPGAS algorithm. We then introduce the new GPGAS algorithm in Section \ref{method}, before demonstrating the performance of the proposed algorithm, compared to the traditional PGAS algorithm, on the challenging regime-switching SSMs in Section \ref{experiments}. Finally, we discuss the proposed method and future avenues for research in Section \ref{discussion}.

\section{Particle Gibbs}\label{background}

We focus on particle Gibbs algorithms, which were proposed by \cite{Andrieu_2010} and have emerged as a popular approach to MCMC targeting the joint distribution of the latent states and model parameters \citep{Chopin_2015, Wigren_2019}. Central to particle Gibbs algorithms are SMC methods, thus we initially introduce SMC and the associated notation.

\subsection{Sequential Monte Carlo}

SMC methods \citep{Gordon_1993} approximate the conditional posterior distribution of the latent states, $p(x_{1:T} \vert y_{1:T}, \theta)$, using importance sampling sequentially targeting each $p(x_{1:t} \vert y_{1:t}, \theta)$ until time $t=T$. The sequential steps are derived by noting that, for general SSMs of the form given in Equation (\ref{eq:SSM}), $p(x_{1:t} \vert y_{1:t}, \theta)$ can be written recursively as

\begin{align}
    p(x_{1:t} \vert y_{1:t}, \theta) &= \frac{p(x_{1:t-1}  \vert y_{1:t-1}, \theta)p(x_t, y_t \vert x_{t-1}, \theta)}{p(y_t \vert y_{1:t-1}, \theta)}, \hspace{2mm} t=1, \dots, T,\label{eq:decomp}
\end{align}

\newpage 
\noindent where, for $t=1$, $p(x_{1:t-1} \vert y_{1:t-1}, \theta)=1$, $p(x_t, y_t \vert x_{t-1}, \theta)=p(x_t, y_t \vert \theta)$, and $p(y_t \vert y_{1:t-1}, \theta)=p(y_1 \vert \theta)$. Thus, suppose we have an importance sampling approximation of $p(x_{1:t-1}  \vert y_{1:t-1}, \theta)$ at the previous time point, given by
\begin{equation*}
    \widehat{p}(x_{1:t-1}  \vert y_{1:t-1}, \theta) = \sum_{m=1}^M W_{1:t-1}(x_{1:t-1}^m) \delta_{x_{1:t-1}^m}(x_{1:t-1}),
\end{equation*}
for a set of $M$ samples (`particles') and associated normalized importance weights, \linebreak $\{x_{1:t-1}^m, W_{1:t-1}(x_{1:t-1}^m)\}_{m=1}^M$, and where $\delta_{x_{1:t-1}^m}(x_{1:t-1})$ denotes the Dirac function at $x_{1:t-1}^m$. We extend this approximation of the conditional distribution at time $t$ via a low-dimensional importance density of the form $q(x_t \vert y_t, x_{t-1}, \theta)$. First, the particles are propagated to time $t$ by sampling a set of particles from the importance distribution, i.e., $x_t^m \sim q(x_t \vert y_t, x_{t-1}^m)$ for $m=1, \dots, M$. Combined with the sequential decomposition of $p(x_{1:t} \vert y_{1:t}, \theta)$ given in Equation (\ref{eq:decomp}), we obtain the approximation:
\begin{align}
    &\widehat{p}(x_{1:t} \vert y_{1:t}, \theta) = \sum_{m=1}^M W_{1:t}(x_{1:t}^m) \delta_{x_{1:t}^m}(x_{1:t}), \nonumber \\ 
    &W_{1:t}(x_{1:t}^m) = \frac{w_{1:t}(x_{1:t}^m)}{\sum_{k=1}^m w_{1:t}(x_{1:t}^k)}, \hspace{2mm}  w_{1:t}(x_{1:t}^m) \propto w_{1:t-1}(x_{1:t-1}^m) \frac{p(x_t^m \vert x_{t-1}^m, \theta)p(y_t \vert x_t^m, \theta)}{q(x_t^m \vert y_t, x_{t-1}^m)}. \label{eq:smc_weights}
\end{align}
For all time points, $t=1, \dots, T$, $w_{1:t}^{1:M}=\{w_{1:t}(x_{1:t}^m)\}_{m=1}^M$ denotes the unnormalised weights and $W_{1:t}^{1:M}=\{W_{1:t}(x_{1:t}^m)\}_{m=1}^M$ the normalized weights such that $\sum_{m=1}^M W_{1:t}(x_{1:t}^m)=1$. Noting that the particles and weights are defined as a function of the particles and weights at the previous time point, we obtain a recursive approximation of $p(x_{1:t} \vert y_{1:t}, \theta)$ given by the set of particle trajectories and (normalized) weights, $\{x_{1:t}^m, W_{1:t}^m\}_{m=1}^M$.

Particle degeneracy occurs in the SMC algorithm when many particles have low weights, eliminating their effective use for posterior estimation. Moreover, degeneracy is inevitable for almost all particle paths as the number of SMC recursions increase \citep{Doucet_2009}. To prevent particle degeneracy, an SMC algorithm typically incorporates an additional resampling step into its recursions, eliminating particles with low weights and replicating those with high weights. Before the importance sampling step at each time point $t=2, \dots, T$, the particle trajectories are sampled from a distribution conditional on their weights, denoted $r(a_t \vert W_{t-1}^{1:M})$. That is, we sample trajectory indices from  
\begin{equation}
   a_t^m \sim r(a_t \vert W_{1:t-1}^{1:M}), \hspace{2mm} m=1, \dots, M, \label{eq:resampling}
\end{equation}
and set $x_{1:t-1}^m=x_{1:t-1}^{a_t^m}$ for $m=1, \dots, M$. However, resampling reduces the diversity in the particles. Thus, resampling steps are often only executed when they are deemed necessary, for example, resampling when the effective sample size of the particles falls below a certain threshold $\psi$  \citep{Moral_2012}. Throughout this paper, we assume standard multinomial resampling, i.e., that $r(a_t \vert W_{1:t-1}^{1:M})$ is a multinomial distribution with probabilities equal to the normalized weights for each $t = 2, \dots, T$, and resample particles by thresholding based on the effective sample size of the particles. Note that, if we resample the particles at time $t-1$ (sample $a_t^{1:N}$), their weights are now equal:

\begin{equation*}
    W_{1:t-1}(x_{1:t-1}^m) =\frac{1}{M}, \quad m=1, \dots, M.
\end{equation*}

\noindent We present the full SMC algorithm with both the sequential importance sampling and resampling steps in Algorithm \ref{alg:SMC}.

\begin{algorithm}[h]
\setstretch{1.25}
\caption{Sequential Monte Carlo (SMC)}\label{alg:SMC}

\Input Importance distributions conditional on fixed $\theta$, $q(x_1 \vert y_1, \theta)$, $\{q(x_t \vert y_t, x_{t-1}, \theta)\}_{t=2}^T$, a number of iterations $M$. A resampling threshold, $\psi$, based on the effective sample size, ESS.

\BlankLine
\BlankLine

 \For{$m=1, \dots, M$}{ 
 sample $x_1^m \sim q(x_1 \vert y_1, \theta)$}
calculate $w_1^{1:M}$ and $W_1^{1:M}$ \Comment{Equation (\ref{eq:smc_weights})} 
 \BlankLine 

\For{$t=2, \dots, T$}{
\For{$m=1, \dots, M$}{
\If{$\text{ESS}<\psi$}{
sample $a_t^m \sim r(a_t \vert W_{1:t-1}^{1:M})$, set $w^m_{1:t-1} = 1/M$ \Comment{Equation (\ref{eq:resampling}})
} 

\textbf{else} set $a_t^m=m$

\BlankLine

sample $x_t^m \sim q(x_t \vert y_t, x_{t-1}^{a_t^m}, \theta$) 

set $x_{1:t}^m=(x_{1:t-1}^{a_t^m}, x_t^{m})$}
\BlankLine

calculate $w_{1:t}^{1:M}$ and $W_{1:t}^{1:M}$ \Comment{Equation (\ref{eq:smc_weights})}}

\BlankLine

\BlankLine

\Return $\{x_{1:T}^m, W_{1:T}^m\}_{m=1}^M$

\end{algorithm}

\newpage 

\subsection{Particle Gibbs}\label{PG}
The particle Gibbs algorithm uses a variant of the SMC algorithm, the conditional SMC (CSMC) algorithm, to sample values for the latent states. The sampled 
latent states are then used as MCMC proposed values targeting $p(x_{1:T} \vert y_{1:T}, \theta)$. These updates can be used as part of an MCMC algorithm targeting the joint distribution, $p(x_{1:T}, \theta \vert y_{1:T})$.

To describe the particle Gibbs algorithm in detail, we start by defining the CSMC algorithm that is used to propose values for the latent states. At each MCMC iteration, the CSMC algorithm first conditions on the current latent states by fixing a `reference trajectory' to their values. The remaining particles are then sampled via standard SMC steps and all particles are weighted as in Equation (\ref{eq:smc_weights}). Without loss of generality, we assume that the last particle trajectory is the reference trajectory, i.e., $x_{1:T}^M=x_{1:T}^{(s-1)}$ for MCMC iteration $s$ and $M$ particles. However, any trajectory can be chosen as the reference trajectory provided that the same trajectory index is chosen for all time points. The CSMC algorithm is presented in Algorithm \ref{alg:CSMC}.

\begin{algorithm*}[h]
\setstretch{1.25}
\caption{Conditional sequential Monte Carlo (CSMC)}\label{alg:CSMC}
\Input A number of particles, $M$, importance distributions, $q(x_1 \vert y_1, \theta)$, $\{q(x_t \vert y_t, x_{t-1}, \theta)\}_{t=2}^T$, a trajectory of latent states, $x_{1:T}^{(s-1)}$ at MCMC iteration $s$, and known parameters, $\theta$. A resampling threshold, $\psi$, based on the effective sample size, ESS.

\BlankLine

\BlankLine

set $x_1^M= x_1^{(s-1)}$ 

\For{$m=1, \dots, M-1$}{sample $x_1^m \sim q(x_1 \vert y_1, \theta)$}
calculate $w_1^{1:M}$ and $W_1^{1:M}$ \Comment{Equation (\ref{eq:smc_weights})}
\BlankLine

\For{$t=2, \dots, T$}{
set $x_t^M=x_t^{(s-1)}$, $a_t^M=M$

\For{$m=1, \dots, M-1$}{
\If{$\text{ESS} < \psi$}{
sample $a_t^m \sim r(a_t \vert W_{1:t-1}^{1:M})$, set $w_{1:t-1}^m=1/M$ \Comment{Equation (\ref{eq:resampling}})
}

\textbf{else} set $a_t^m=m$

\BlankLine

sample $x_t^m \sim q(x_t \vert y_t, x_{t-1}^{a_t^m}, \theta)$

set $x_{1:t}^m=(x_{1:t-1}^{a_t^m}, x_t^{m})$}

calculate $w_t^{1:M}$ and $W_{1:t}^{1:M}$ \Comment{Equation (\ref{eq:smc_weights})}
}

\BlankLine
\BlankLine

\Return $\{x_{1:T}^m, W_{1:T}^m\}_{m=1}^M$
\end{algorithm*}

Once the CSMC recursions have been completed, the particle Gibbs algorithm proposes MCMC values for the latent states from the resulting approximation of $p(x_{1:T} \vert y_{1:T}, \theta)$, $\{x_{1:T}^m, W_{1:T}^m\}_{m=1}^M$. The proposed values are always accepted, resulting in Gibbs steps. Finally, the model parameters are updated using standard and often low-dimensional Metropolis-Hastings (M-H) or Gibbs steps targeting $p(\theta \vert x_{1:T}, y_{1:T})$. This particle Gibbs algorithm results in MCMC samples converging to the joint distribution $p(x_{1:T}, \theta \vert y_{1:T})$ and is given in Algorithm \ref{alg:PG}.

\begin{algorithm*}[h]
\caption{Particle Gibbs}\label{alg:PG}
\Input A number of particles, $M$, initial values, $x_{1:T}^{(0)}$ and $\theta^{(0)}$, a number of iterations, $S$,
importance distributions, $q(x_1 \vert y_1, \theta)$, $\{q(x_t \vert y_t, x_{t-1}, \theta)\}_{t=2}^T$, a Gibbs or Metropolis-Hastings sampling scheme to update $\theta$ from $p(\theta \vert x_{1:T}, y_{1:T})$. 

\BlankLine

\BlankLine

\For{$s=1, \dots S$}{update $\theta^{(s)}$ from $p(\theta \vert x_{1:T}^{(s-1)}, y_{1:T})$

run Algorithm \ref{alg:CSMC} with $q(x_1 \vert y_1, \theta)$, $\{q(x_t \vert y_t, x_{t-1}, \theta)\}_{t=2}^T$, $x_{1:T}^{(s-1)}$, and $\theta=\theta^{(s)}$

sample $x_{1:T}^{(s)}$ from $\{x_{1:T}^m, W_{1:T}^m\}_{m=1}^M$}

\BlankLine 
\BlankLine

\Return $\{x_{1:T}^{(s)}, \theta^{(s)}\}_{s=1}^S$ approximating $p(x_{1:T}, \theta \vert y_{1:T})$
\end{algorithm*}

The CSMC algorithm ensures the particle Gibbs proposed values not only target the entire state vector but these values are always accepted. \cite{Andrieu_2010} and \cite{Chopin_2015} establish that the particle Gibbs state samples are distributed according to $p(x_{1:T} \vert y_{1:T}, \theta)$ upon convergence. The authors show that the algorithm samples from an extended target distribution that admits $p(x_{1:T} \vert y_{1:T}, \theta)$ as a marginal distribution due to a corrective unbiased estimate of the likelihood term. Thus, the particle Gibbs algorithm converges to $p(x_{1:T} \vert y_{1:T}, \theta)$ but latent state samples are also always `accepted' in the MCMC steps.

\subsection{Particle Gibbs with ancestor sampling}
The mixing of the particle Gibbs algorithm can be poor when sample impoverishment occurs in the CSMC approximation \citep{Chopin_2015, Rainforth_2016, Wigren_2019}. In severe cases of sample impoverishment, the reference trajectory is nearly always proposed (and accepted) in the particle Gibbs steps since it is fixed. The MCMC algorithm therefore remains at the same values for the latent states for many iterations, leading to poor mixing. To improve the mixing of particle Gibbs methods, \cite{Lindsten_2014} proposed the particle Gibbs with ancestor sampling (PGAS) algorithm, which uses CSMC with ancestor sampling (CSMC-AS) to artificially recompose the particle Gibbs reference trajectory, ensuring that unique values for the latent states are proposed at each MCMC iteration.

The CSMC-AS algorithm recomposes the reference trajectory at each CSMC forward recursion by artificially re-assigning its particle history. We re-assign the particle history by first noting that in the CSMC algorithm, the reference trajectory at each time $t = 2, \dots, T$ is indexed by $a_t^M$. Thus, to recreate the history of the reference trajectory, the CSMC-AS algorithm samples new values for $a_t^M$ at each time $t$. New values for $a_t^M$ are sampled according to the probability that the associated trajectory generated the reference particle, denoted $\tilde{w}_t^m$ for time $t = 2, \dots, T$, and given by 
\begin{equation}
    \tilde{w}_t^m \propto w_{1:t-1}^m p(x_t^{(s-1)} \vert x_{t-1}^m, \theta), \hspace{2mm} m=1, \dots, M, \label{eq:anc_weights}
\end{equation}
where $x_t^{(s-1)}$ denotes the reference particle (state sample at iteration $(s-1)$) at time $t$. The weights are normalized so that they sum to one, giving normalized weights $\tilde{W}_t^m = \tilde{w}_t^m/\sum_{k=1}^M \tilde{w}_t^k$ and the new ancestor is sampled using these weights, i.e., $a_t^M$ is sampled from $\{m, \tilde{W}_t^m\}_{m=1}^M$ at each time $t$. Finally, the reference trajectory at time $t = 2, \dots, T$ is recreated by attaching the current reference particle to its new likely history, $x_{1:t}^M=(x_{1:t-1}^{a_t^M}, x_t^{(s-1)})$. We summarize the CSMC-AS algorithm in Algorithm \ref{alg:CSMCAS}.

\begin{algorithm*}[h]
\setstretch{1.25}
\caption{Conditional sequential Monte Carlo with ancestor sampling (CSMC-AS)}\label{alg:CSMCAS}
\Input A number of particles, $M$, importance distributions, $q(x_1 \vert y_1, \theta)$, $\{q(x_t \vert y_t, x_{t-1}, \theta)\}_{t=2}^T$, a trajectory of latent states, $x_{1:T}^{(s-1)}$ at MCMC iteration s, and known parameters, $\theta$. A resampling threshold, $\psi$, based on the effective sample size, ESS.

\BlankLine

\BlankLine

set $x_1^M= x_1^{(s-1)}$ 

\For{$m=1, \dots, M-1$}{sample $x_1^m \sim q(x_1 \vert y_1, \theta)$}
calculate $w_1^{1:M}$ and $W_1^{1:M}$ \Comment{Equation} (\ref{eq:smc_weights})
\BlankLine

\For{$t=2, \dots, T$}{
set $x_t^M=x_t^{(s-1)}$ 

\For{$m=1, \dots, M-1$}{
\If{$\text{ESS}<\psi$}{
sample $a_t^m \sim r(a_t \vert W_{t-1}^{1:M})$, set $w_{1:t-1}^m=1/M$  \Comment{Equation (\ref{eq:resampling})}
} 

\textbf{else} set $a_t^m=m$

\BlankLine

sample $x_t^m \sim q(x_t \vert y_t, x_{t-1}^{a_t^m}, \theta)$}

\vspace{2ex}
calculate $\tilde{W}_t^{1:M}$ \Comment{ancestor sampling, Equation (\ref{eq:anc_weights})}

sample $a_t^M$ from $\{m, \tilde{W}_t^m\}_{m=1}^M$

\vspace{1ex}
set $x_{1:t}^m=(x_{1:t-1}^{a_t^m}, x_t^{m})$, $m=1, \dots, M$

calculate $w_{1:t}^{1:M}$ and $W_{1:t}^{1:M}$ \Comment{Equation (\ref{eq:smc_weights})}
}

\BlankLine

\BlankLine

\Return $\{x_{1:T}^m, W_{1:T}^m\}_{m=1}^M$

\end{algorithm*}

Once the CSMC-AS recursions have been completed, the PGAS algorithm samples new values for the latent states $x_{1:T}^{(s)}$ at iteration $s$, targeting $p(x_{1:T} \vert y_{1:T}, \theta)$ and the parameters are updated targeting $p(\theta \vert y_{1:T}, x_{1:T})$. The full PGAS algorithm is given in Algorithm \ref{alg:PGAS}.

\begin{algorithm*}[h]
\caption{Particle Gibbs with ancestor sampling (PGAS)}\label{alg:PGAS}
\Input A number of particles, $M$, initial values $x_{1:T}^{(0)}$, $\theta^{(0)}$, a number of iterations $S$, 
importance distributions, $q(x_1 \vert y_1, \theta)$, $\{q(x_t \vert y_t, x_{t-1}, \theta)\}_{t=2}^T$, 
a Gibbs or Metropolis-Hastings sampling scheme to update $\theta$ from $p(\theta \vert x_{1:T}, y_{1:T})$. 

\BlankLine

\BlankLine

\For{$s=1, \dots S$}{update $\theta^{(s)}$ from $p(\theta \vert x_{1:T}^{(s-1)}, y_{1:T})$

run Algorithm \ref{alg:CSMCAS} with $q(x_1 \vert y_1, \theta)$, $\{q(x_t \vert y_t, x_{t-1}, \theta)\}_{t=2}^T$, $x_{1:T}^{(s-1)}$, and $\theta=\theta^{(s)}$

sample $x_{1:T}^{(s)}$ from $\{x_{1:T}^m, W_{1:T}^m\}_{m=1}^M$}

\BlankLine

\BlankLine

\Return $\{x_{1:T}^{(s)}, \theta^{(s)}\}_{s=1}^S$ approximating $p(x_{1:T}, \theta \vert y_{1:T})$
\end{algorithm*}

\subsubsection{Optimal importance distributions}

PGAS methods are shown to improve upon the mixing of particle Gibbs algorithms both theoretically and in a wide range of examples \citep{Berntorp_2017, Chopin_2015, Nonejad_2015, Wigren_2019}. However, PGAS methods can still be inefficient when there is a high rate of sample impoverishment in the SMC algorithm. If sample impoverishment is particularly prevalent, the pool of trajectories at each CSMC recursion may represent the posterior distribution poorly and the MCMC sampler may not explore the space sufficiently even if the history of the reference trajectory is recomposed in ancestor sampling steps \citep{Rainforth_2016}. 

\newpage 
The mixing of particle Gibbs methods can be improved by simulating particles in high posterior regions, tackling the initial sample impoverishment problem. Ideally, the SMC importance distributions generate particles that exactly represent the posterior distribution, thus producing uniformly distributed weights and maximizing the number of particles that survive the resampling steps of the CSMC-AS algorithm. As in Equation (\ref{eq:smc_weights}), the SMC \linebreak \newpage \noindent steps initially sample particles from the importance distribution, $x_t^m \sim q(x_t \vert y_t, x_{t-1}^m, \theta)$, $m=1, \dots, M$, and then approximates the conditional distribution of the latent states by
\begin{align}
    &\widehat{p}(x_{1:t} \vert y_{1:t}, \theta) = \sum_{m=1}^M W_{1:t}^m \delta_{x_{1:t}^m}(x_{1:t}), \nonumber \\ 
    &W_t^m = \frac{w_{1:t}^m}{\sum_{k=1}^M w_{1:t}^k}, \quad   w_{1:t}^m \propto w_{1:t-1}^m \frac{p(x_t^m \vert x_{t-1}^m, \theta)p(y_t \vert x_t^m, \theta)}{q(x_t^m \vert y_t, x_{t-1}^m)}, \label{eq:SMC_2}
\end{align}
for $t=1, \dots, T$, where $w_{1:t}^{1:M}=\{w_{1:t}^m\}_{m=1}^M$ and $W_{1:t}^{1:M}=\{W_{1:t}^m\}_{m=1}^M$ denote the set of unnormalised weights and normalized $w_t^{1:M}$ weights at time $t$, respectively. To minimize sample impoverishment at each recursion (i.e., produce uniformly distributed weights), the optimal approach samples particles from $p(x_{1:t} \vert y_{1:t}, \theta)$ directly. This approach also approximates the likelihood of all previous observations \citep{Branchini_2021, Chopin_2020, Elvira_2019}. An alternative approach is to sample particles in a locally-optimal manner and sample from the target distribution at each time point, $p(x_t \vert y_t, x_{t-1}, \theta)$. This results in new multiplicative weight terms at each time point (Equation (\ref{eq:SMC_2})) that are uniformly distributed. This is typically referred to as the `optimal' importance distribution \citep{Doucet_2009} but is a function of the current observation and does not admit a tractable sampling distribution for general SSMs. 

Several approaches have been proposed to approximate the optimal importance distributions, including via Gaussian approximations of the given SSM \citep{Andrieu_2003}, deterministic optimization-based approximations in annealing schemes \citep{Donnet_2017}, and variational approximations of the posterior \citep{He_2022}. A popular approach is the auxiliary particle filter \citep{Carpenter_2000, Pitt_1999, Pitt_2001} which approximates the optimal importance distribution for general SSMs. At each resampling step of the SMC recursions, the auxiliary particle filter accounts for the current observation, often via a simulated approximation of the optimal importance distribution \citep{Elvira_2018}. However, since the CSMC steps of a particle Gibbs algorithm are simply used to formulate proposal distributions for the latent states, the computational cost associated with the use of the auxiliary particle filter within each CSMC sweep of the MCMC algorithm can accumulate quickly. We propose novel optimal-type importance distributions using discrete HMM approximations to the SSM, which also reduce computational cost in the particle Gibbs iterations and produce a computationally efficient approach.

\section{Grid particle Gibbs with ancestor sampling}\label{method}
In this section, we introduce the proposed GPGAS algorithm. For general SSMs, the optimal PGAS importance densities are not available in closed form. We therefore propose general-use importance densities that use a tractable HMM approximation of the SSM. In Step 1, we present the approximate HMM construction (following a similar approach to \cite{Llewellyn_2023}) and point mass filtering \citep{Bucy_1971, Kitagawa_1987, Langrock_2012, deValphine_2002, Matousek_2019}. In Step 2, we introduce the novel tractable discrete approximations of the optimal importance distribution at each time point. The approximations of the optimal importance distributions are then used within the CSMC-AS steps of the PGAS algorithm.

\subsection{Step 1: Approximate HMM}
\begin{figure}[h]
    \centering
    \includegraphics[width=\textwidth]{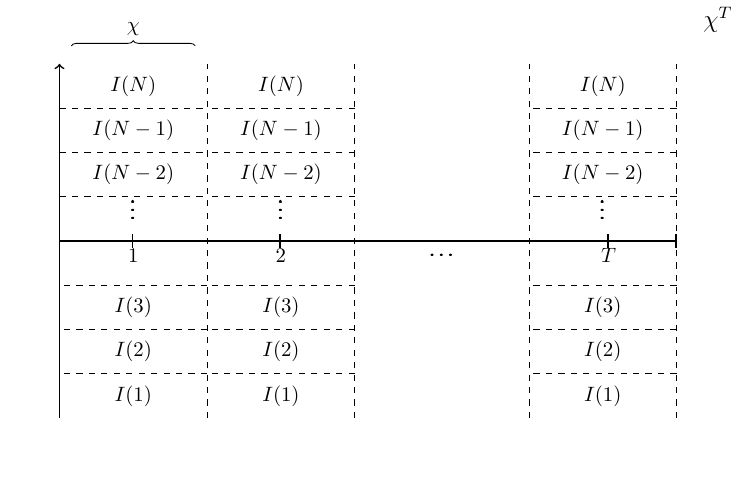}
    \caption{Partition of the state space into equally-sized grid cells, the same for each time point. The grid cells are labeled by the interval they cover.}
    \label{fig:grid}
\end{figure}

We present the algorithm for one-dimensional state spaces and note that extensions to higher-dimensional spaces are possible (this is discussed further in Sections \ref{experiments} and \ref{discussion}). To approximate the SSM by a deterministic HMM, the state space is first partitioned into grid cells. That is, at each time point, we partition the state space, $\chi$, into $N$ intervals that span the space with no overlap. The intervals form grid cells when the state space is partitioned for all time points. See Figure \ref{fig:grid} for a graphical representation of the partition. For notational simplicity, we assume that the grid cells are the same for all time points and denote them by $I(n)$, $n=1, \dots, N$, but this can be easily relaxed. 

We assume the non-infinite grid cells (Figure \ref{fig:grid}) are equally sized (i.e., the grid cells cover the same amount of the state space at each time point). Additional grid cell definitions are described by \cite{Llewellyn_2023, Matousek_2019}. However, we note that one should consider the trade-off between computational cost and efficiency. This is discussed further in Section \ref{discussion}. 
 
The grid cell indices, $\{1, \dots, N\}$, can be interpreted as the discrete states of an HMM, with dynamics defined by the SSM \citep{Kitagawa_1987, Langrock_2011,Langrock_2013}. Since we define the same grid cells for all time points, $t=1,\dots,T$, the HMM state transition probabilities are also the same for all time points. Letting $B_t$ denote the random variable of the grid cell indices at time $t$, we define the HMM for $k, n \in \{1, \dots, N\}$ as follows:
\vspace{2mm}

\noindent{\mbox{\emph{Initial state probabilities:}}}
\begin{equation*}
  P(B_1=n \vert \theta) = \int_{I(n)} p(x_1 \vert \theta) dx_1.  
\end{equation*}
\noindent{\mbox{\emph{State transition probabilities}:}}
\begin{equation*}
   P(B_t = n \vert B_{t-1}=k, \theta)=\int_{I(n)} \int_{I(k)} p(x_t \vert x_{t-1}, \theta) dx_{t-1} dx_t, \hspace{2mm} \text{for all } t=2, \dots, T.  
\end{equation*}
\noindent{\mbox{\emph{Observed state distribution}:}}
\begin{equation}
    p(y_t \vert B_t=n, \theta)=\int_{I(n)} p(y_t \vert x_t, \theta) dx_t, \hspace{2mm} t=1, \dots, T.\label{eq:HMM_full}     
\end{equation}

\noindent In general, these HMM probabilities do not admit a closed-form expression. Thus, we apply a deterministic midpoint integration approach to approximate the HMM (as in \citealp{Llewellyn_2023, Newman_1998}). Let the length and midpoint of the $n^{th}$ interval, $I(n)$, be denoted by $L(n)$ and $\xi(n)$ respectively. We define the length of and midpoint in infinite cells arbitrarily (Section \ref{practical}; \citealp{Llewellyn_2023}) by, for example, defining the length at the average length of the finite cells and setting the midpoint equal to the midpoint of this finite grid cell. We approximate the HMM in Equation (\ref{eq:HMM_full}) by
\begin{align}
    \widehat{P}(B_1 = n \vert \theta) &\propto L(n) p(\xi(n) \vert \theta), \nonumber \\ 
    \widehat{P}(B_t=n \vert B_{t-1}=k, \theta) &\propto L(n)  L(k)  p(\xi(n) \vert \xi(k), \theta), \hspace{2mm} \text{for all} \hspace{2mm} t=2, \dots, T, \nonumber \\ 
    \widehat{p}(y_t \vert B_t=n, \theta) &\propto L(n) p(y_t \vert \xi(n), \theta), \hspace{2mm} t=1, \dots, T, \label{Eq:approx_HMM}
\end{align}
\noindent for grid cells indices $k, n=1, \dots, N$. Each of these probabilities are bounded to ensure \linebreak \newpage \noindent that they are non-zero, and normalized over the grid cells so that they sum to one for each $t=1, \dots, T$. 

\subsection{Step 2: HMM-based importance distributions}
We use the HMM approximation to formulate SMC importance distributions to improve particle distribution and sample impoverishment. We start by defining the following discrete approximation of the optimal importance distribution using the approximate HMM:
\begin{align*}
    \widehat{P}(B_1 = n \vert y_1, \theta) &\propto \widehat{P}(B_1 = n \vert \theta)\widehat{p}(y_1 \vert B_1=n, \theta), \nonumber \\ 
    \widehat{P}(B_t = n \vert y_t, B_{t-1} = k, \theta) &\propto \widehat{P}(B_t = n \vert B_{t-1}=k, \theta) \widehat{p}(y_t \vert B_t = n, \theta), \hspace{2mm} t=2, \dots, T,
\end{align*}
for grid cells indices $k, n=1, \dots, N$. At time $t$, we sample $M$ grid cell indices (one for each SMC particle trajectory) from the associated discrete approximation. That is, at $t=1$, we sample $b_1^m$ for $m=1, \dots, M$ from 
\begin{equation*}
    \{n, \widehat{P}(B_1 = n \vert y_1, \theta)\}_{n=1}^N.
\end{equation*}
At time $t =2, \dots, T$, we sample $b_t^m$ from
\begin{equation*}
    \{n, \widehat{P}(B_t=n \vert y_t, B_{t-1}=b_{t-1}^m, \theta)\}_{n=1}^N,
\end{equation*}
for each $m=1, \dots, M$. Given a set of sampled grid cell indices at time $t$, $b_t^{1:M}$, we propose continuously-valued particles by sampling from within the grid cells associated with these indices. We therefore define continuous importance distributions over the space of each grid cell. We assume that the importance distributions within each grid cell are defined independently of $\theta$ and the data, i.e., the importance distributions are of the form $q(x_t \vert B_t=n)=q(x_t \vert x_t \in I(n))$, for $n=1, \dots, N$. Examples of such within-cell distributions include uniform distributions for bounded grid cells and truncated Gaussian distributions for infinite grid cells.

To sample a particle at each time $t$, a grid cell is first sampled from the approximate optimal importance distribution conditional on the grid cell at the previous time point. A continuous particle value for time $t$ is then sampled from within the sampled grid cell for time $t$, resulting in a sampled grid cell index and particle, $b_1^m$ and $x_1^m$ respectively. When repeated for the specified number of particles, we obtain a set of grid cells and particles at time $t$, $\{b_t^m, x_t^m\}_{m=1}^M$, from the importance distributions:
\begin{align}
    q(x_1, B_1 \vert y_1, \theta) &= \widehat{P}(B_1 \vert y_1, \theta) q(x_1 \vert B_1), \nonumber \\ 
    q(x_t, B_t \vert y_t, B_{t-1}, \theta) &= \widehat{P}(B_t \vert y_t, B_{t-1}, \theta)q(x_t \vert B_t), \quad \text{for $t=2, \dots, T$}.
    \label{eq:full_imp}
\end{align}
In addition, we have that \mbox{$q(x_1, B_1 = b_1 \vert y_1, \theta)=q(x_1 \vert y_1, \theta)$} and $q(x_t, B_t=b_t \vert y_t, B_{t-1}=b_{t-1}, \theta)=q(x_t \vert y_t, B_{t-1}=b_{t-1}, \theta)$ for all $t=2, \dots, T$ since we sample particles such that $x_t \in I(b_t)$ for all $t$. These distributions are defined over the state space since the grid cells and within-cell distributions assign non-zero probability everywhere in the space.

\bigskip

We note that the HMM transition probability approximations are time invariant. That is, the HMM transition probability approximation at time $t=2$ also applies at times $t=3, \dots, T$. A time-dependent HMM approximation can be derived. For example, the exact values of the particles at the previous time point could be used to approximate the transition probabilities. Specifically, we may formulate a proposal distribution of the form $q(x_t, B_t \vert y_t, x_{t-1}, \theta)$ (replacing Equation (\ref{eq:full_imp})) using a transition matrix of the form $\widehat{P}(B_t \vert y_t, x_{t-1}, \theta)$ (replacing Equation (\ref{Eq:approx_HMM})). Although this may improve the accuracy of the HMM approximation, additional transition probability calculations are required (for each unique particle and each time point). Thus, the potentially improved mixing properties need to be balanced with the additional computational cost.  

\subsection{Grid importance distribution within particle Gibbs with ancestor sampling}
Within the CSMC-AS steps of the PGAS algorithm, the GPGAS algorithm samples grid cells and particles according to Equation (\ref{eq:full_imp}), denoted $\{b_t^m, x_t^m\}_{m=1}^M$ for $t=1, \dots, T$. Thus, the SMC approximation of $p(x_{1:t} \mid y_{1:t}, \theta)$ under the proposed importance distribution is given by 

\begin{align*}
    \widehat{p}(x_{1:t} \vert y_{1:t}, \theta) &= \sum_{m=1}^M W_{1:t}^m \delta_{x_{1:t}^m}(x_{1:t}),\nonumber \\[2ex]
    w_1^m &\propto \frac{p(x_1^m \vert \theta)p(y_1 \vert x_1^m, \theta) }{q(x_1^m, B_1=b_1^m \vert y_t, \theta)},  \nonumber \\[1ex]
    w_{1:t}^m &\propto w_{1:t-1}^m  \frac{ p(x_t^m \vert x_{t-1}^m, \theta)p(y_t \vert x_{t}^m, \theta) }{q(x_t^m, B_t=b_t^m \vert y_t, x_{t-1}^m, \theta)}, \quad t=2, \dots, T, 
\end{align*}
where $W_{1:t}^m=w_{1:t}^m/\sum_{k=1}^M w_{1:t}^k$ is the weight associated with both $b_{1:t}^m$ and $x_{1:t}^m$, and is defined recursively. To use the proposed importance distribution within a CSMC-AS algorithm, we simply replace the importance distributions and weights in Algorithm \ref{alg:PGAS} with those defined above, resulting in the GPGAS algorithm. We present this version of the GPGAS algorithm in Algorithm \ref{alg:gridpf}.

\begin{algorithm*}[!h]
\caption{Grid particle Gibbs with ancestor sampling (GPGAS)}\label{alg:gridpf}
\Input A number of particles, $M$, and grid cells, $N$. A grid with indices $B_{1:T}$ and importance distributions $\{q(x_t \vert x_t \in I(n))\}_{n=1}^N$, for all $t=1, \dots, T$. Initial values $x_{1:T}^{(0)}$, $\theta^{(0)}$ and number of iterations $S$, and a Gibbs or Metropolis-Hastings sampling scheme to update $\theta$ from $p(\theta \vert x_{1:T}, y_{1:T})$.

\BlankLine

\BlankLine

\For{$s=1, \dots S$}{update $\theta^{(s)}$ from $p(\theta \vert x_{1:T}^{(s-1)}, y_{1:T})$
\BlankLine
\BlankLine

\textbf{Approximate HMM (Step 1)}

calculate $\widehat{P}(y_1 \vert B_1=n, \theta^{(s)})$, $\widehat{P}(B_1=n \vert \theta^{(s)})$, $n=1, \dots, N$ 
 
calculate $\widehat{P}(B_t = n \vert B_{t-1} = k, \theta^{(s)})$, $k, n=1, \dots, N$

\For{$t=2, \dots, T$}{
calculate $\widehat{P}(y_t \vert B_t=n, \theta^{(s)})$, $n=1, \dots, N$
}

\BlankLine 
\BlankLine
\textbf{Formulate importance distributions (Step 2)}

calculate $\widehat{P}(B_1=n \vert y_1, \theta^{(s)})$, $n=1, \dots, N$

\For{$t=2, \dots, T$}{
calculate $\widehat{P}(B_t = n \vert y_t, B_{t-1} = k, \theta^{(s)})$, $k, n=1, \dots, N$  
}

\BlankLine
\BlankLine

Run a PGAS step (Algorithm \ref{alg:CSMCAS}) with $M$ particles, $\theta=\theta^{(s)}$, $x_{t-1}^{(s-1)}$, and importance distributions $q(x_1, B_1 \vert y_1, \theta^{(s)})$, $\{q(x_t, B_t \vert y_t, B_{t-1}, \theta^{(s)})\}_{t=2}^T$ defined in Equation (\ref{eq:full_imp})

\BlankLine

sample $x_{1:T}^{(s)}$ from $\{x_{1:T}^m, W_T^m\}_{m=1}^M$}

\BlankLine

\BlankLine

\Return $\{x_{1:T}^{(s)}, \theta^{(s)}\}_{s=1}^S$ approximating $p(x_{1:T}, \theta \vert y_{1:T})$

\end{algorithm*}

\newpage 
\subsection{Computational and practical considerations}\label{practical}
In this section, we describe the associated computational and practical aspects of the GPGAS algorithm that influence its efficiency. We first note some important computational strategies that can be implemented universally, independent of the model considered:

\begin{itemize}
    \item [1.] As illustrated in Algorithm \ref{alg:gridpf}, only one approximate HMM transition probability matrix needs to be calculated per MCMC iteration since the grid cells are the same for all time points.
    \item [2.] The discrete approximations to the optimal importance distributions at time $t\geq 2$, only need to be calculated for grid cells containing particles at the previous time point. This computational strategy reduces the computational cost of each GPGAS iteration from $\mathcal{O}(N^2T)$ to $\mathcal{O}(N^2 + N\sum_{t=2}^T \tilde{N}_{t-1})$, where $\tilde{N}_{t-1}$ denotes the number of grid cells containing particles at time $t-1$ of the GPGAS iteration.
    \item [3.] Given the sampled grid cells at time $t$, many of the particles at time $t$ are identically distributed according to the importance distributions within each grid cell. Thus, we can sample multiple particles from the same importance distribution simultaneously to reduce computational cost.
\end{itemize}

\noindent Aside from the computational adjustments that can be made universally, there are model-dependent practical considerations, particularly with respect to how the grid cells are defined. We provide general guidance in relation to these. We note that for the examples considered in Section \ref{experiments}, performance was robust within these general guidelines, and efficient decisions were made in relation to each point, where appropriate, using pilot tuning over a small number of MCMC iterations.

\begin{itemize}
    \item [a)] The overall computational cost of the HMM approximation can be reduced by fixing the approximation after a given number of iterations. We consider this a sensible approach assuming that the HMM approximation is stable when calculated using the average HMM approximation past a certain number of iterations, $\tilde{s}$. In this paper, we simply fix the HMM approximation using the parameter mean estimates of several samples after a given number of iterations, $\tilde{s}$, and note that it may be possible to obtain an improved approximation by fixing the HMM probabilities using the average HMM probability estimates after the given iterations. In either case, the value of $\tilde{s}$ should be chosen to balance the reduction in computational cost with the accuracy of the importance distributions.         
    \item [b)] To ensure that any value in the state-space with reasonable posterior mass can be proposed, the majority of finite grid cells should be set to ensure that areas of the state-space with significant probability are covered. Excessively large ranges should be avoided to ensure that computational cost is not spent in effectively zero-density areas of the state space. 
    \item [c)] We sample particles within each grid cell using standard (and computationally inexpensive) distributions, for example, uniform distributions in the finite grid cells and truncated Gaussian distributions in the outer (infinite) grid cells. In the implementations of Section \ref{experiments}, we parameterized the truncated Gaussian distributions by setting their mean equal to the `midpoint' of the associated grid cell, defined at a distance from the finite boundary equal to the distance between the midpoints and finite boundaries in the finite grid cells.
\end{itemize}

\section{State-space models with regime switching}\label{experiments} 
We investigate the performance of the GPGAS algorithm when applied to the challenging case of SSMs with regime switching. Regime-switching SSMs allow the observation or transition models of an SSM to change abruptly between a discrete set of `regimes'. At each time point, the choice of regime determines the observation and transition model, and the regime label is assumed to be first-order Markovian, forming an additional unobserved latent process. Despite the several well-known applications of regime-switching SSMs, including tracking maneuvering targets \citep{Karlsson_2000, Barshalom_2002, Liang-qun_2009} and modeling economic and financial data \citep{Hamilton_1989, Kim_1999, Fruhwirth_2001, Kim_2022}, computational methods for fitting the latent states and model parameters of a regime-switching SSM can be inefficient. Since the SSM is non-linear, a natural approach often applied is particle Gibbs sampling. However, standard SMC steps within the particle Gibbs algorithm are known to degenerate when the state switches, requiring many particles and a high computational cost to combat sample impoverishment \citep{Doucet_2001, Driessen_2005}.

\newpage 
Various strategies have been proposed to merge deterministic techniques with importance sampling to combat sample impoverishment for regime-switching SSMs, including the deterministic allocation of particles to regimes heuristically or using posterior model probability approximations \citep{El-Laham_2022, Martino_2017, Urteaga_2016}. The GPGAS algorithm is intuitively similar to these approaches but provides an approach to joint state and parameter inference.

We investigate the performance of the proposed GPGAS algorithm when applied to two classes of regime-switching models. In the first example, we focus on a simulated stochastic volatility model with regime-switching to investigate the performance of the GPGAS algorithm relative to several current efficient approaches: the PGAS algorithm using both the bootstrap and auxiliary particle filters \citep{Lindsten_2014, Pitt_1999}, and the PMPMH algorithm proposed by \cite{Llewellyn_2023}. Further, we also explore the performance of each method under two different model parameterizations that are known to impact the efficiency of traditional SMC methods, understanding some of the settings in which each algorithm can be applied efficiently. The second example applies the best-performing algorithms in this initial example to a challenging real-world regime-switching model for COVID-era tourism demand in Edinburgh. We demonstrate that the GPGAS algorithm provides a practical and efficient method even for this challenging example.

\subsection{Stochastic volatility with leverage}\label{SVL}
We initially focus on the model for stochastic volatility described by \cite{So_1998} and \cite{Kim_2015b}. In this model, a two-state regime process, denoted $s_{1:T}=(s_1, \dots, s_T)$, $s_t \in \{1, 2\}$ for each $t$, captures switching in the level of U.S. stock market log volatility over time. Transitions from the first and second regimes occur with probability $\pi_{12}$ and $\pi_{21}$ respectively, and the regime labels, $s_{1:T}$, each correspond to a parameter, $\gamma_1$ or $\gamma_2$. The level of the latent log volatility process, $x_{1:T}$, is a function of these parameters and determines the variance of the observations, $y_{1:T}$. Mathematically, this stochastic volatility SSM with regime switching can be written as follows:
\bigskip

\noindent{\mbox{\emph{State transition distribution:}}}
\begin{equation*}
    x_t = \gamma_{s_t} + \phi (x_{t-1} - \gamma_{s_{t-1}}) + \eta_t, \quad \eta_t \sim N(0, \sigma^2_{\eta}).
\end{equation*}
\noindent{\mbox{\emph{Observed state distribution:}}}
\begin{equation*}
    y_t = \exp \left( \frac{x_{t-1}}{2} \right) \epsilon_t, \quad \epsilon_t \sim N(0, 1).
\end{equation*}
\noindent{\mbox{\emph{Regime transition probabilities:}}}
\begin{equation*}
    P(s_t=j \mid s_{t-1}=i) = \pi_{ij}, \quad i,j \in \{1, 2\}, \label{SV}
\end{equation*}
for $t=1, \dots, T$ where $\phi$ is an autoregressive scaling parameter and $\sigma^2_{\eta}>0$ is the system process variance. In addition, the initial continuous latent state is defined as $x_0=\mu$, the initial regime label is $s_0=1$, and the expected duration in each regime is specified to be the same, i.e, $\pi_{22}=\pi_{11}$. We consider that $x_{1:T}$, $s_{1:T}$, and the set of model parameters, $\theta=(\gamma_1, \gamma_2, \phi, \sigma^2_{\eta}, \mu,  \pi_{11})$, are unknown.

The duration of the regimes (persistence) is determined by $\pi_{11}$ and can influence how well an SMC algorithm approximates the posterior distribution of the latent states. In general, degeneracy rates increase when the true state switches. Thus, increasing the number of states that switch generally increases degeneracy rates and reduces the accuracy of the SMC approximation of the posterior distribution of the latent states. We therefore investigate the performance of the proposed GPGAS algorithm under different levels of regime persistence, resulting in different sets of simulated data: $y_{1:T}^{(1)}$, simulated using $\pi_{11}=0.85$, and $y_{1:T}^{(2)}$, using $\pi_{11}=0.95$ (reducing the number of state switches). Each data set is simulated using $T=500$ time points and model parameters $\theta = (\gamma_1, \gamma_2, \phi, \sigma^2_{\eta}, \mu, \pi_{11}) = (-5, 5, 0.95, 0.1, 1, \pi_{11})$. We present each set of simulated data in Figure \ref{fig:SV_data} and the associated prior distributions and sampling schemes are given in Appendix A.

\begin{figure}[h]
    \centering
    \begin{subfigure}[t]{0.495\textwidth}
         \centering
    \includegraphics[width=0.35\paperwidth]{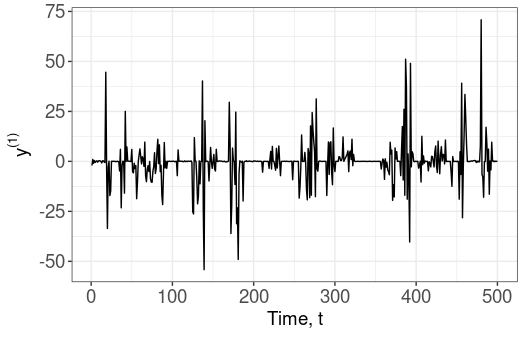}
    \caption{}
    \end{subfigure}
    \begin{subfigure}[t]{0.495\textwidth}
         \centering
     \includegraphics[width=0.35\paperwidth]{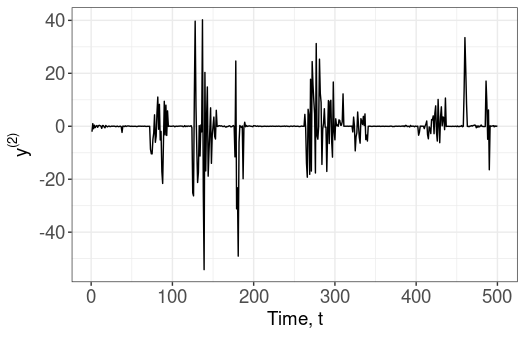}
     \caption{}
    \end{subfigure}
    \caption{Simulated data from the stochastic volatility model: $y^{(1)}_{1:T}$ using $\pi_{11}=0.85,$ $T=500$, and $y^{(2)}_{1:T}$ using $\pi_{11}=0.95,$ $T=500$.}
    \label{fig:SV_data}
\end{figure}

\newpage 

\subsubsection{Computational decisions} 

We specify the computational and practical decisions to implement the GPGAS algorithm for this example with reference to Section \ref{practical}. We update the latent states, $x_{1:T}$ and $s_{1:T}$, from their conditional distribution, $p(x_{1:T}, s_{1:T} \vert y_{1:T}, \theta)$. The joint latent state process distribution is $p(x_t, s_t \vert x_{t-1}, s_{t-1}, \theta)$. The observed state distribution is given by $p(y_t \vert x_t, \theta)$ since the observations only depend on the continuous latent state $x_t$. We apply an HMM approximation to these densities, using the exact (transition) probabilities for $s_{1:T}$ in the joint HMM transition probability calculations since these states are discrete. The following relates to the grid cells used in the space of the continuous states, $\chi$, to approximate $p(x_t  \vert x_{t-1}, s_t, s_{t-1}, \theta)$ and $p(y_t \vert x_t, \theta)$.

The GPGAS algorithm is implemented using the computational strategies in points (1-3) of Section \ref{practical}. The practical choices with respect to points (a-c) in Section \ref{practical} are as follows:
\begin{itemize}
    \item [a)] In all implementations, we fix the HMM approximations using the posterior mean of each parameter estimated using $1000$ samples after iteration $\tilde{s}=2000$. These values were found to be reasonable from short pilot tuning runs.
    \item [b)] Using these pilot tuning runs, we establish that a range of $[-12, 12]$ for the finite grid cells ensures that any value in the state space with reasonable posterior mass can be proposed. 
    \item[c)] To sample within each grid cell, we use uniform and truncated Gaussian distributions as described in Section \ref{practical}, and note that the truncated Gaussian distributions have variance $2.4$ ($10\%$ of the finite grid cell range). 
\end{itemize}

\subsubsection{Results}\label{SV:results}
We present the results for the GPGAS, PGAS, PGAS with the auxiliary particle filter, and PMPMH algorithms with various numbers of grid cells and particles. A resampling threshold ($\psi$) of $25\%$ of the effective sample size in the SMC recursions is universally favorable for both the GPGAS and PGAS algorithms in this case. Each implementation is executed $10$ times for $10{,}000$ iterations on one core and a $1.6$ GHz CPU and we compare the performance of each implementation to `ground truth' runs. These ground truth runs consist of the PGAS algorithm with $M=5000$ particles, taking around $89$ hours to complete $10{,}000$ iterations under both sets of simulated data, $y_{1:T}^{(1)}$ and $y_{1:T}^{(2)}$. 

\begin{figure}[H]
    \centering
    \includegraphics[width=0.9\textwidth]{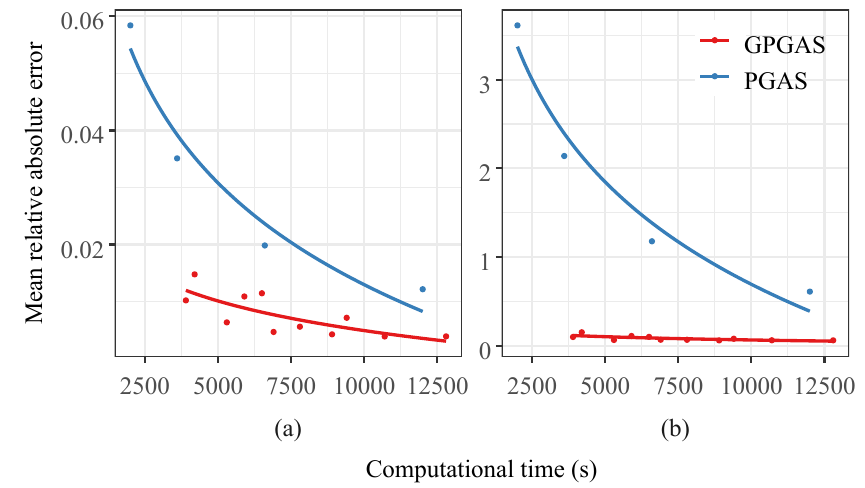}
    \caption{Mean relative absolute errors versus computational time for the (a) posterior mean and (b) posterior variance estimates of the continuous latent states, $x_{1:T}$, under $y^{(1)}_{1:T}$ ($\pi_{11}=0.85$). Each point represents a different combination of $N\in \{10, 25, 50, 100\}$ grid cells and $M \in \{10, 25, 50, 100, 200\}$ particles. Non-convergent implementations are excluded. Computational time is the time in seconds to complete the $10000$ iterations.}\label{fig:SV1}
\end{figure}

\begin{figure}[H]
    \centering
    \includegraphics[width=0.9\textwidth]{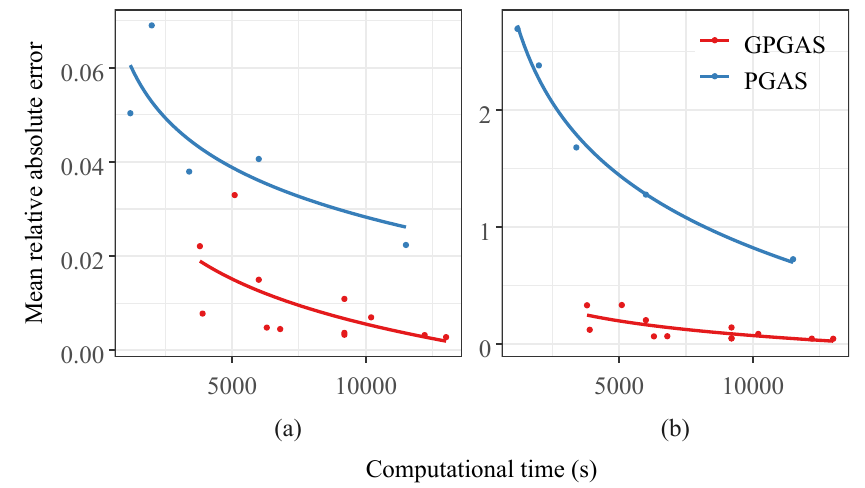}
    \caption{Mean relative absolute errors versus computational time for the (a) posterior mean and (b) posterior variance estimates of the continuous latent states, $x_{1:T}$, under $y^{(2)}_{1:T}$ ($\pi_{11}=0.95$). Each point represents a different combination of $N\in \{10, 25, 50, 100\}$ grid cells and $M \in \{10, 25, 50, 100, 200\}$ particles. Non-convergent implementations are excluded. Computational time is the time in seconds to complete the $10000$ iterations.}\label{fig:SV2}
\end{figure}

The auxiliary particle filter implementation uses simulation of the auxiliary weights and requires a large computational cost for sufficiently accurate approximation of the auxiliary weights to prevent sample impoverishment: at least $30$ particles to approximate each auxiliary weight and $50$ particles in the CSMC-AS recursions, taking around $3$ hours to reach errors comparable to the cheapest standard PGAS implementation. Similarly, the PMPMH algorithm requires the calculation of many transition matrices, and a large computational cost, to achieve comparable errors in the posterior estimates. Thus, we focus the remaining results on the PGAS algorithm and proposed GPGAS algorithm and present the results in Figures \ref{fig:SV1} and \ref{fig:SV2}.

The GPGAS algorithm leads to improved posterior mean and variance estimates for a fixed computational time, with notable improvements in the mean relative absolute errors. The algorithm scales well with both the number of grid cells and the number of particles for the regime-switching implementations considered in this section. This demonstrates the potential gains in efficiency from combining deterministic approximations of the SMC importance distributions with the computational strategies presented in points 1-3 of Section \ref{practical}. 

The reason for the differences in performance between the algorithms is likely related to the associated degeneracy rates. For approximately equivalent run times, the GPGAS algorithm reduces the average number of states not updated in the SMC steps by $11-50\%$ depending on the implementation. To see the effect of regime switching on the sample impoverishment and accuracy of posterior estimates, we present additional results in Appendix B. The results summarize the performance of each algorithm according to switching and non-switching states and demonstrate the improved efficiency and robustness of the GPGAS algorithm to estimate both the switching and non-switching states.

\subsection{Tourism demand regime-switching state-space model}\label{tourism}
\begin{figure}[h]
\centering
\begin{subfigure}{0.45\textwidth}
    \includegraphics[width=\textwidth]{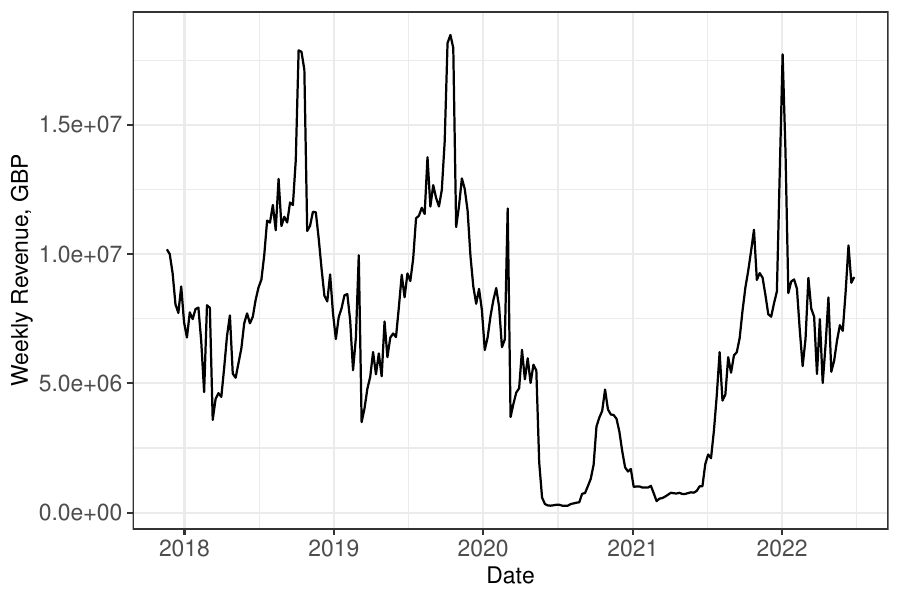}
    \caption{}
    \label{plot:revenue_all}
\end{subfigure}
\hfill
\begin{subfigure}{0.45\textwidth}
    \includegraphics[width=\textwidth]{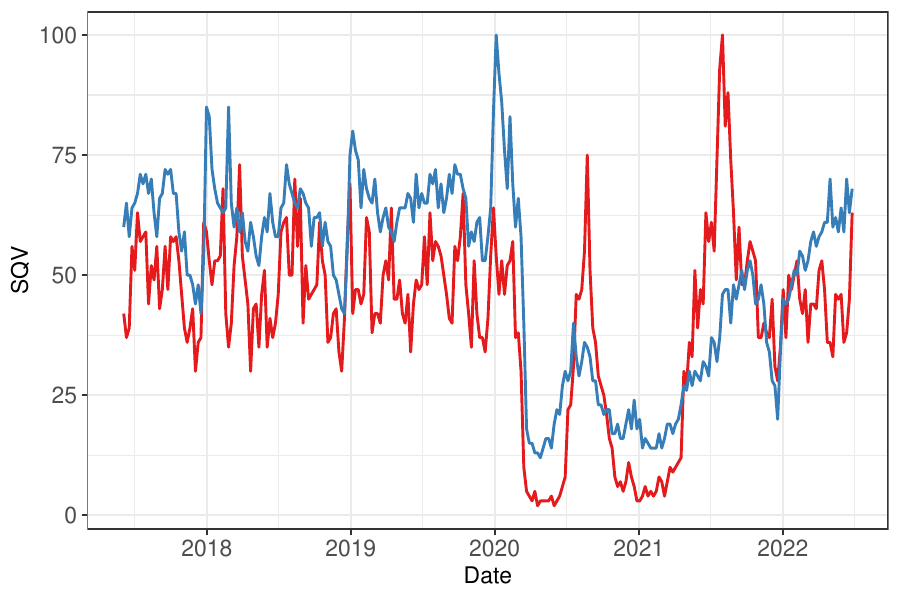}
    \caption{}
    \label{plot:trends}
\end{subfigure}
\caption{Plots of (a) the aggregate weekly revenue of hotels in Edinburgh in Great British Pounds (GBP) and (b) the data for two Google search query volumes (SQVs): UK searches `things to do in Edinburgh' in red and Global searches for `flights to Edinburgh' in blue.}
\end{figure}

We consider a challenging real data regime-switching example motivated by the impact of the COVID-19 pandemic on Edinburgh's tourism industry, the biggest contributor to Scottish tourism revenue before COVID \citep{LGT_2018}. In post-COVID recovery plans, understanding the nature of recovery is essential for business communities and policymakers to formulate appropriate policy responses \citep{Lawrence_2020, OECD_2020}. As in \cite{Llewellyn_2023b}, we consider response data measuring weekly aggregate hotel revenue (a proxy for tourism demand) of over $300$ hotels in Edinburgh pre and post-COVID (Figure \ref{plot:revenue_all}). We also consider additional covariate data from Google Trends, comprised of $254$ weekly search query volumes aiming to capture behavioral responses to the pandemic in the absence of systematic patterns. Example series from this data set are provided in Figure \ref{plot:trends}.

The hotel revenue data are modeled as a regime-switching SSM with a library of structural components and shrinkage priors to capture dynamic model uncertainty in the COVID period. We denote the hotel revenue data up to time $T$ by $y_{1:T}=(y_1, \dots, y_T)$ and model these data via structural time series components (trend and seasonality). We include additional covariates derived from the Google Trends data, $\mathbf{G}_{1:T}=(g_{1:T}^1, \dots, g_{1:T}^{254})$, consisting of principal components that reduce the dimension of the data set \citep{Bishop_1998}. The parameterization of the model varies depending on the regime, giving the regime-switching SSM for the tourism demand data for $t=1, \dots, T$:

\begin{align*}
       y_t \vert \mu_t &\sim \text{logNormal}(\lambda_t^{s_t} + \mu_t -a_{s_t}(PC_t - u_t), \sigma^2_{\epsilon_{s_t}}), \\
       \mathbf{G}_t &\sim N(\mathbf{W}_{s_t} PC_t, \sigma^2_{\eta_{s_t}}), \\
        \mu_t \vert \mu_{t-1} &\sim N(\mu_{t-1} + b_{s_t}, \sigma^2_{\mu_{s_t}}), \\
        u_t \vert u_{t-1} &\sim N(u_{t-1} + c_{s_t}, \sigma^2_{u_{s_t}}),  \\ 
        P(s_t=j &\vert s_{t-1}=i)=\pi_{ij}, \quad i,j \in \{1, 2\},
\end{align*}
where $\lambda_t^{i} \in \{\lambda_1^i, \dots, \lambda_{52}^i\}$, $i \in \{1, 2 \}$ and $\mu_{1:T}$ are regime-dependent annual seasonal components and trend terms respectively. The Google Trends data relate to the principal components, $PC_{1:T}=(PC_1, \dots, PC_T)$, via a $254$-dimensional vector of weights, $\mathbf{W}_{s_t}$ for each $t=1, \dots, T$, with variance $\sigma^2_{\eta_{s_t}}>0$ and relate to the tourism demand data with trend and a linear regression parameters, denoted by $u_{1:T}$ and $a_{1:2}$ respectively. The observation process variance is given by $\sigma^2_{\epsilon_{s_t}}>0$. The trend terms are assumed to follow linear auto-regressive processes with $\mu_0$ and $u_0$ unknown parameters and system process variances $\sigma^2_{\mu_{s_t}}$ and $\sigma^2_{u_{s_t}} >0$. Finally, the parameterization of the model can vary according to the regime label at each time point $s_{1:T}=(s_1, \dots, s_T)$. We assume that the initial regime is arbitrarily set to $s_0=1$ and note that the unknown parameters are $\theta=(\lambda_{1:T}^{1:2}, PC_{1:T}, a_{1:2}, \sigma^2_{\epsilon_{1:2}}, \mathbf{W}_{1:2}, \sigma^2_{\eta_{1:2}}, b_{1:2}, \sigma^2_{\mu_{1:2}},$ $c_{1:2}, \sigma^2_{u_{1:2}}, \pi_{11}, \pi_{22}, \mu_0, u_0)$. We assign independent priors to each parameter, provided in Appendix \ref{moddev} along with the sampling schemes.

\subsubsection{Computational decisions} 
We describe the computational GPGAS approach to inferring the latent states $(\mu_{1:T}, u_{1:T}, s_{1:T})$ and model parameters, $\theta$. To reduce the computational cost associated with defining grid cells in a three-dimensional latent space, we first sample $(s_{1:T}, \mu_{1:T})$ jointly from their full conditional distribution, followed by $u_{1:T}$ from its full conditional distribution. In both cases, the GPGAS algorithm is implemented following the computational strategies in points (1-3) of Section \ref{practical}. We use the exact HMM transition probabilities in the discrete regime label space. In the continuous spaces of $\mu_{1:T}$ and $u_{1:T}$, we make the following decisions with respect to points (a-c) (Section \ref{practical}): 
\begin{itemize}
    \item [a)] The HMM approximations are fixed after iteration $\tilde{s}=1500$ using the posterior mean of each parameter in iterations $1000-1500$. This is a lower value than Section \ref{SVL} to address the computational cost associated with HMM approximation in two continuous state dimensions, and was found to be reasonable via pilot tuning.
    
    \item [b)] Using this pilot tuning run, the finite grid cells were found to cover a reasonable posterior mass over ranges $[-5,20]$ in the state space of each $\mu_t$ and $[-300, 1000]$ in the state space of each $u_t$.
    
    \item[c)] As in Section \ref{SVL}, we sample from within each grid cell using uniform distributions over the finite cells, and otherwise we sample values for $\mu_t$ and $u_t$ using truncated Gaussian distributions with variances $2.5$ and $130$ respectively ($10\%$ of the range of the finite grid cells). 
\end{itemize}

\subsubsection{Results}\label{tourism:results}
We compare the performance of the GPGAS algorithm with the PGAS algorithm. Due to the large range of the finite cells required to update $u_{1:T}$, and the accuracy of the HMM approximation required, we present an additional approach that updates $(s_{1:T}, \mu_{1:T})$ using the GPGAS algorithm and $u_{1:T}$ using the PGAS algorithm (referred to as \textit{GPGAS + PGAS}). For a fair comparison, we assess the performance of these approaches when compared to a PGAS algorithm using the same conditional structure for the updates (\textit{conditional PGAS}), as well as a PGAS algorithm jointly updating $(s_{1:T}, \mu_{1:T}, u_{1:T})$ at each iteration (\textit{joint PGAS}). The SMC resampling threshold for all algorithms is set at the case-optimal level of $\psi=50\%$ of the effective sample size, and we test the efficiency of each algorithm using $M=10, 25, 50, 100, 200, 300, 400$ particles and $N=25, 50, 100, 200, 300, 400$ grid cells.

We execute each implementation (combination of tuning parameters) $10$ times for 1 hour on one core and a $1.6$ GHz CPU and compare the results to a joint updating PGAS algorithm with $M=5000$ particles (the `ground truth'). Each ground truth run takes around $97$ hours to complete $25{,}000$ iterations. The results for the most efficient implementations of each algorithm are presented in Table \ref{tab:4_1} and are defined as those achieving the lowest mean squared errors compared to the ground truth. Since there are several model parameters and three latent state processes, we summarize each approach by the errors in posterior predictive estimates, i.e., those estimated from samples from the marginal distribution $p(\tilde{y}_{1:T} \vert y_{1:T})$ for new observations $\tilde{y}_{1:T}$.

\newpage 

\begin{table}[H]
    \centering
    \begin{tabular}{ccccccc}
    \hline 
        & & & \multicolumn{2}{c}{\textbf{MRAE}} & & \textbf{Iterations}  \\ 
        & \textbf{ESS} & \textbf{Mean} & \textbf{Var} & \textbf{50\% CrI} & \textbf{90\% CrI} & \textbf{per hour}  \\ 
            \hline
       \textbf{Joint PGAS} & 3100 & 0.040 & 0.476 & 0.050 & 0.104 & 15000 \\
       \textbf{Conditional PGAS} & 3300 & 0.039 & 0.376 & 0.048 & 0.097 & 11900 \\ 
        \textbf{GPGAS} & 1100 & 0.044 & 0.368 & 0.068 & 0.156 & 9500   \\
        \textbf{GPGAS + PGAS} & 5800 & 0.029 & 0.261 & 0.041 & 0.089 & 16100 \\
        \hline 
        \end{tabular}
    \caption{Effective sample size (ESS), mean relative absolute error (MRAE) of the estimated posterior predictive mean and variance (Var), and average RRMSE of equal-tailed credible intervals (CrI), and the number of iterations completed within $1$ hour (Iterations per hour). Shown for the most efficient implementations of each algorithm: joint and conditional PGAS with $200$ particles, and the GPGAS and GPGAS + PGAS algorithms with $200$ grid cells and $100$ particles for both the GPGAS and PGAS updates.}
    \label{tab:4_1}
\end{table}

\noindent Overall, the results indicate that the GPGAS updates of $\mu_{1:T}$ and $s_{1:T}$ and PGAS updates of $u_{1:T}$ (GPGAS + PGAS) is the most efficient approach. The GPGAS-only algorithm is less efficient than the GPGAS and PGAS combined approach due to the large high posterior density range requiring many grid cells to achieve reasonable HMM approximation error in the updates for $u_{1:T}$. However, the GPGAS algorithm appears to improve efficiency at switching points, increasing the number of unique particles at these points by $5-7\%$ on average, and thus provides an efficient approach for updating $\mu_{1:T}$ and $s_{1:T}$.

\section{Discussion}\label{discussion}
We present an efficient particle Gibbs approach to fitting general SSMs using a deterministic grid within the SMC steps. We show that this GPGAS approach improves efficiency for challenging regime-switching SSMs where current SMC-based approaches are inefficient due to sample impoverishment. By combining a deterministic grid with SMC steps, we have utilized grid-based approaches and their ability to direct particles to areas of high posterior mass while reducing their overall computational cost and improving their scalability in the number of grid cells, and the scalability of SMC steps in the number of particles. Further, the SMC corrections have reduced the number of tuning parameters associated with current grid-based approaches (for example in \citealp{Llewellyn_2023}), and their sensitivity, improving their practical use.

The combination of deterministic grid and SMC methods presents a number of interesting points for future research. To further reduce the computational cost of the method, one possibility is to introduce a deterministic grid on the space of the observations, thereby reducing the number of observed state probability matrix calculations in the HMM approximations. It may also be possible to reduce computational cost whilst retaining mixing properties by adapting the number of grid cells at each time point, reducing the number of grid cells when there is little uncertainty in the latent states. However, any such adaptations of the GPGAS algorithm should be made considering potentially reduced mixing properties.

The computational time of the GPGAS algorithm may also be reduced in real terms by parallelization. As with other SMC approaches, trajectories of particles can be sampled in parallel. A particularly efficient approach could group parallel computations by the grid cells containing particles from the previous time point, thus avoiding the additional computational cost from relaxing computational strategy 2 of Section \ref{practical}. Further approaches to parallelization can also be considered and are discussed, for example, in \cite{Verge_2015}. Note that, as with any parallelized algorithm, the computational cost associated with re-synchronization should also be considered \citep{Henriksen_2012}. 

In this paper, we explored the combination of PGAS and GPGAS updates to improve efficiency. In Section \ref{tourism} in particular, we show that the equally-sized grid cell GPGAS algorithm can have a high computational cost when applied to states with a large high posterior density range. It may be possible to improve the efficiency of the proposed algorithm in such cases using a state-centered or similar approach (for example in \citealp{Llewellyn_2023}), provided that this still provides a valid particle Gibbs algorithm. The grid cell boundaries could vary through time according to the empirical quantiles of the particles at each time point or the current states in the MCMC iterations. However, the equally-sized grid cells of the GPGAS algorithm scale well with the state dimension, requiring few transition matrix calculations in the MCMC steps. Therefore, approaches that improve the HMM model approximation for large high posterior mass ranges whilst maintaining a small number of transition matrix calculations could be explored. A possible approach could define the grid cells in the same way for all or several time points, setting the grid cells according to coarsely-approximated quantiles of the true posterior distribution via, for example, variational Bayes approximations \citep{Onizuka_2023}. However, the computational gains should be balanced with the computational cost of the chosen approach. Further, such approaches may depend highly on the current states and perform poorly if the HMM approximation is fixed in future iterations to reduce computational cost.

An additional consideration is the design of grid cells on high-dimensional spaces, which is often non-trivial \citep{Smidl_2013, Dunik_2019} and is a particular challenge when it is inefficient to sample lower-dimensional state dimensions conditional on other state dimensions. One interesting idea would involve combining the grid-based approach and standard SMC importance distributions within the SMC steps, applying the grid-based importance distribution only to state dimensions that are likely to degenerate. Other approaches may include projecting the grid definition to lower-dimensional spaces \citep{Tidefelt_2009}. This is a challenging and active area for future research.

Finally, the proposed grid-based importance distribution could be extended to other SMC-based methods. In particular, the grid importance distribution could be applied to improve sample impoverishment in filtering applications with fixed model parameters. In this case, the grid-based approach does not require multiple transition and observed state probability matrix approximations (across iterations) and is thus computationally inexpensive. However, for online parameter inference, using for example the nested particle filter \citep{Crisan_2018, PerezVieites_2021}, the method may be computationally costly, requiring many transition and observation probability matrix approximations for different model parameter values. One possibility would be to calculate HMM approximations for groups of similar model parameter samples, reducing the number of HMM approximations required. This presents a particularly interesting avenue for future research, extending the grid importance distribution to other SMC-based methods to combat sample impoverishment efficiently.

\bibliographystyle{apalike} 

\begin{appendix}
\section{Parameter prior distributions and sampling schemes for the stochastic volatility model}
The (independent) priors for the unknown parameters of Section 4.1.2, $\theta = (\gamma_1, \gamma_2, \phi, \sigma^2_{\eta}, \mu, \pi_{11})$, are given for both data sets by:
\begin{align}
\gamma_1 &\sim N(-5, 10), \nonumber \\
\gamma_2 &\sim N(5, 10), \nonumber \\ 
\phi &\sim N(0.95, 1), \nonumber \\
\sigma^2_{\eta} &\sim \text{InvGamma}(2.01, 0.101), \nonumber \\ 
\mu &\sim N(1, 1),  \nonumber\\ 
\pi_{11} &\sim \text{Beta}(9.9875, 1.7625),
\end{align}
where InvGamma denotes an inverse gamma distribution and the Gaussian distributions are parameterized by their variance. Each unknown model parameter is sampled in the same way for each model parameterization using conditional Gibbs updates.

\newpage

\section{Results by switching/non-switching states}

\begin{figure}[h]
    \centering
    \includegraphics[width=\textwidth]{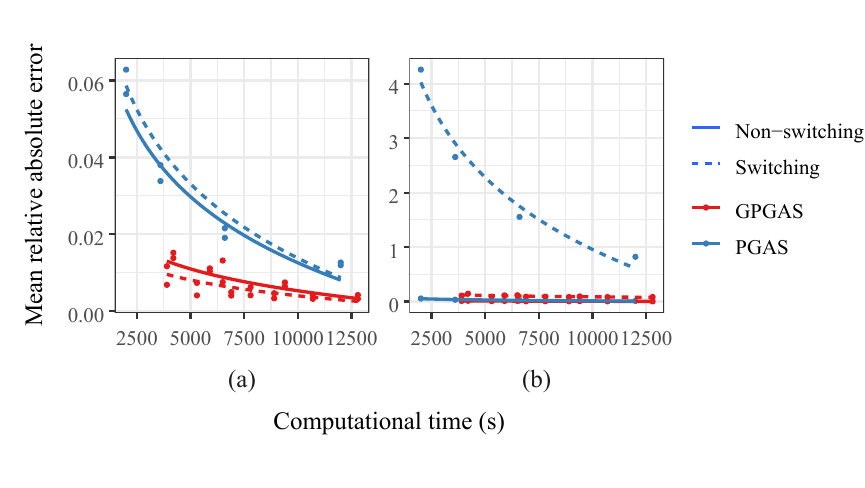} 
    \caption{Mean relative absolute errors for the (a) posterior mean and (b) posterior variance estimates by non-switching and switching states with computational time for $y^{(1)}_{1:T}$ (simulated with $\pi_{11}=0.85$). Each point represents a different combination of $N\in \{10, 25, 50, 100\}$ grid cells and $M \in \{10, 25, 50, 100, 200\}$ particles; non-convergent implementations are excluded. Computational time is measured as the time in seconds taken to complete the $10000$ iterations.}\label{fig:SV1_cps}
\end{figure}

\begin{figure}[H]
    \centering
    \includegraphics[width=\textwidth]{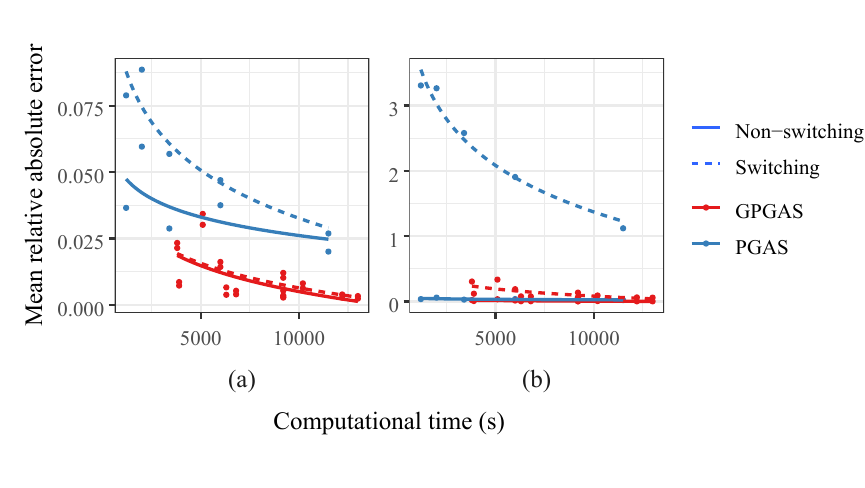}
    \caption{Mean relative absolute errors for the  (a) posterior mean and (b) posterior variance estimates by non-switching and switching states with computational time for $y^{(2)}_{1:T}$ (simulated with $\pi_{11}=0.95$). Each point represents a different combination of $N\in \{10, 25, 50, 100\}$ grid cells and $M \in \{10, 25, 50, 100, 200\}$ particles; non-convergent implementations are excluded. Computational time is measured as the time, in seconds (s), taken to complete the $10000$ iterations.}\label{fig:SV2_cps}
\end{figure}

We present additional results to support those in Section \ref{SV:results}, showing the change in the relative root mean squared error according to whether the states switch. Figure \ref{fig:SV1_cps} shows the the results according to switching/non-switching states for the first data set considered in Section \ref{SV:results}, $y_{1:T}^{(1)}$ with $\pi_{11}=0.85$, and Figure \ref{fig:SV2_cps} shows the the results according to switching/non-switching states for the second data set, $y_{1:T}^{(2)}$ with $\pi_{11}=0.95$. The results in the figures demonstrate that the GPGAS algorithm is comparatively robust to switching in the states, with comparable errors in both the mean and variance errors when allowing for Monte Carlo error.

\section{Parameter prior distributions and sampling schemes for the tourism demand model}\label{moddev}
To specify the tourism demand model in Section \ref{tourism}, we assign (independent) priors to the unknown model parameters:
\begin{align}
    &\lambda_t^1 \sim N(16, 0.5), \quad t = 1, \dots, 52, \nonumber \\
    &\lambda_t^2 \sim N(0, 1), \quad t = 1, \dots, 52, \nonumber \\
    &PC_t, a_i, b_i, c_i, \mu_0, u_0 \sim N(0, 1), \quad t=1, \dots, T, \quad i=1, 2,  \nonumber \\
    &W_i^k \sim N(0, 1), \quad i=1, 2, \quad k=1, \dots, 254, \nonumber \\
    &\sigma^2_{\epsilon_i}, \sigma^2_{\eta_i}, \sigma^2_{\mu_i}, \sigma^2_{u_i} \sim \text{InvGamma}(2, 1), \quad i=1, 2, \nonumber \\
    &\pi_{ii} \sim \text{Beta}(9.9875, 1.7625), \quad i=1, 2.
\end{align}
We note that we apply simple zero-centered priors (ridge priors) for many parameters to avoid over-fitting. The Gaussian distributions are parameterized by their variance. The choice of non-zero-centered priors for $\lambda_{1:52}^1$  corresponds to the prior knowledge that seasonality is present in at least one period (for example, pre-COVID). The prior parameters for $\lambda_{1:52}^1$ are chosen to reflect the assumption that average weekly revenue is in the order of $1 \times 10^7$ (hence the log average revenue is around 16). We also assume persistent regimes via the priors for $\pi_{11}$ and $\pi_{22}$, which have expected values of $0.85$ and variances of $0.01$.

These priors give conditional Gibbs updates for $b_{1:2}$, $c_{1:2}$, $\mu_0$, $u_0$, $\mathbf{W}_{1:2}$, $\sigma_{\eta_{1:T}}$, $\sigma_{\mu_{1:T}}$, $\sigma_{u_{1:T}}$, $\pi_{11}$, and $\pi_{22}$. The remaining parameters are independently sampled from Gaussian random walk proposal distributions: the $\lambda_{1:52}^{1}$ with variance 0.15, the $\lambda_{1:52}^{2}$ with variance 1, the $PC_{1:T}$ with variance $0.01$, $a_1$ with variance $1 \times 10^{-6}$, the $a_{2}$ with variance $1 \times 10^{-4}$, the $\sigma^2_{\epsilon_1}$ with variance $5 \times 10^{-4}$, and the $\sigma^2_{\epsilon_2}$ with variance $1 \times 10^{-2}$.

\end{appendix}

\end{document}